\documentclass[aps,pra,twocolumn,showpacs, superscriptaddress,10pt,groupedaddress]{revtex4-2}

\usepackage{amsmath,braket,amssymb,natbib, graphicx,amsthm, xcolor,outlines, mathrsfs,multirow,hhline}
\usepackage{multirow}
\usepackage{mathtools}
\usepackage{bm}
\usepackage{dcolumn}

\usepackage{dsfont}

\usepackage[english]{babel}
\usepackage[utf8x]{inputenc}
\usepackage[T1]{fontenc}
\usepackage{listings}
\usepackage[colorinlistoftodos]{todonotes}
\usepackage{hyperref}
\usepackage{cleveref}
\begin{document}
\title{Compactifying linear optical unitaries using multiport beamsplitters}

\author{P. A. Ameen Yasir}
\email{apooliab@uni-mainz.de}
\affiliation{Institute of Physics, Johannes-Gutenberg University of Mainz,  Staudingerweg 7, 55128 Mainz, Germany}
\author{Peter van Loock}
\email{loock@uni-mainz.de}
\affiliation{Institute of Physics, Johannes-Gutenberg University of Mainz,  Staudingerweg 7, 55128 Mainz, Germany}

\date{\today}

\begin{abstract}
We show that any $N$-dimensional unitary matrix can be realized using a finite sequence of concatenated identical fixed multiport beamsplitters (MBSs) and phase shifters\,(PSs). Our construction is based on a Lie group theorem applied to existing decompositions. Using the Bell-Walmsley-Clements framework, we prove that any $N$-dimensional unitary requires $N+2$ phase masks, $N-1$ fixed MBSs, and $N-1$ BSs. Our scheme requires only $\mathcal{O}(N)$ fixed, identical components (MBSs and BSs) compared to the $\mathcal{O}(N^2)$ fixed BSs required by conventional schemes (e.g., Clements), all while keeping the same number of PSs. Experimentally, these MBS can be realized as a monolithic element via femtosecond laser writing, offering superior performance through reduced insertion losses. As an application, we present a reconfigurable linear optical circuit that implements a three-dimensional unitary emerging in the unambiguous discrimination of two nonorthogonal qubit states. 
\end{abstract}

\maketitle

%\tableofcontents

%%%%%%%%%%%%%%%%%%%%%%%%%%%%%%%%%%%%%%%%%%%%%%%%%%%%%%%%%%%%%%%%%%%%%%%%%%%%%%%%%%%%%%%%%%%%%%%%%%%%%%%%%
\section{Introduction} \label{int}

Linear optical quantum computing (LOQC) is a model of quantum computation that employs photons as information carriers and utilizes linear optical elements---such as beamsplitters (BSs), phase shifters (PSs), and photodetectors---to implement quantum operations~\cite{knill2001,kok2007}. In this framework, quantum information is encoded in the quantum states of photons, with two widely used encoding schemes: polarization encoding and multi-rail (or path) encoding~\cite{kok2007, kok2010}. Polarization encoding is restricted to the two-dimensional Hilbert space defined by horizontal and vertical polarization states; unitary operations in this space can be realized using sequences of quarter- and half-wave plates~\cite{simon90}. In contrast, path encoding allows, in principle, the implementation of arbitrary linear transformations across any number of modes. Beyond circuit-based LOQC, alternative models such as measurement-based quantum computation~\cite{briegel2009} and fusion-based quantum computation~\cite{bartolucci2023} offer promising routes to scalable photonic quantum computing. 

Reck~\textit{et al.}\cite{reck94} introduced the first reconfigurable linear optical setup capable of realizing any $N$-dimensional unitary operator, $U_N$, using a sequence of BSs and PSs. Building on this, Clements~\textit{et al.}\cite{clements2016} proposed a rectangular architecture that reduced the overall optical path length of the interferometer, enhancing its scalability and robustness. Bell and Walmsley~\cite{bell2021} introduced a universal interferometer with symmetric Mach-Zehnder interferometers\,(sMZI) and external PSs that resulted in reduced propagation loss. Saygin~\textit{et al.}\cite{saygin2020} explored an alternative decomposition based on static multichannel blocks in place of BSs. However, they were unable to prove the universality of their approach, even in the three-dimensional case. More recently, Arends~\textit{et al.}~\cite{arends2024} studied the decomposition of high-dimensional unitaries into multimode blocks of arbitrary dimension, demonstrating that such decompositions become increasingly loss-tolerant as the block size grows. Numerous other reconfigurable linear optical architectures have also been proposed~\cite{guise2018, pai2019, su2019, tan2019, fldzhyan2020, ortiz2021, hamerly2022a, hamerly2022b,zhou2024}  over the years, further advancing the field. 

An alternative approach employs cascaded discrete Fourier transform\,(DFT) blocks interleaved with phase masks, though the number of layers required for universal unitary synthesis remains under investigation. It was conjectured that $N+1$ or $N+2$ phase masks would suffice for an $N$-mode transformation~\cite{saygin2020,pereira2025}. Pastor~\textit{et al.}~\cite{pastor2021} later derived an upper bound of $6N+1$ phase masks, which was significantly tightened to $2N+5$ in Ref.~\cite{girouard2025}.

On the experimental front, a six-mode interferometer based on the Reck scheme was demonstrated in Ref.~\cite{carolan2015}. QuiX reported a 20-mode universal quantum photonic processor employing the Clements scheme~\cite{taballione2023}. Quite recently, Quandela has demonstrated a 12-mode reconfigurable photonic integrated circuit using the Clements scheme, along with on-chip boson sampling of six photons~\cite{maring2024}.

This paper approaches the unitary realization problem from a distinct perspective. Recent technological advances indicate that the mass production of multiport BSs (MBSs) ---also known as $N$-splitters~\cite{zukowski97,loock2002} --- is likely to become feasible in the near future~\cite{carine2020}. Motivated by this, we first prove that any unitary operator $U_N$ can be constructed using a finite sequence of concatenated {\it identical} fixed MBSs, leveraging a Lie group decomposition theorem. We show that any $N$-dimensional unitary can be implemented using $N+2$ phase masks, $N-1$ fixed MBSs, and $N-1$ BSs within the Bell-Walmsley-Clements (BWC) interferometer framework~\cite{bell2021}. Crucially, this component count results in a fixed element complexity of $\mathcal{O}(N)$, offering a substantial advantage over established methods like the Clements scheme, which requires $\mathcal{O}(N^2)$ fixed BSs, all while maintaining the same number of required tunable PSs. We then propose a reconfigurable linear optical setup for $U_3$, implemented using four fixed tritters. As an application, we demonstrate a three-dimensional reconfigurable interferometer capable of distinguishing two non-orthogonal qubit states. 

The paper is structured as follows. Section\,\ref{mbs} introduces BS and describes how an MBS can be constructed from them. Section\,\ref{lo} briefly reviews standard linear optical decompositions. Section\,\ref{fi} presents the realization of arbitrary unitaries using identical MBSs and BSs within the BWC interferometer. Section\,\ref{cr} concludes with final remarks.

\section{Multiport Beamsplitters} \label{mbs}

%We first talk about symmetric and asymmetric Mach-Zehnder interferometers\,(MZI) and extend the notion to tritters and MBSs in this Section. Then we mention that how a tritter can be experimentally implemented as a monolithic device
We first discuss symmetric and asymmetric Mach-Zehnder interferometers (MZIs). We then extend this concept to include three-port devices, such as tritters and MBSs, within this Section. Finally, we detail the experimental implementation of a tritter as a monolithic device. An MBS is a natural generalization of the standard two-port BS~\cite{zukowski97,loock2002}, implementing a linear transformation in multiple optical modes. MBSs have been employed in a variety of quantum optics applications, including the generation of entangled states---such as Greenberger-Horne-Zeilinger (GHZ) and W states~\cite{lim2005_BS,bhatti2023,kumar2023,loock2000,aoki2003}---generalized Hong-Ou-Mandel (HOM) interference experiments~\cite{lim2005_HOM}, multiphoton quantum interference~\cite{huang2025}, and quantum random number generation~\cite{carine2020}.

\begin{figure}[tbp]
\centering
\includegraphics[scale=0.6]{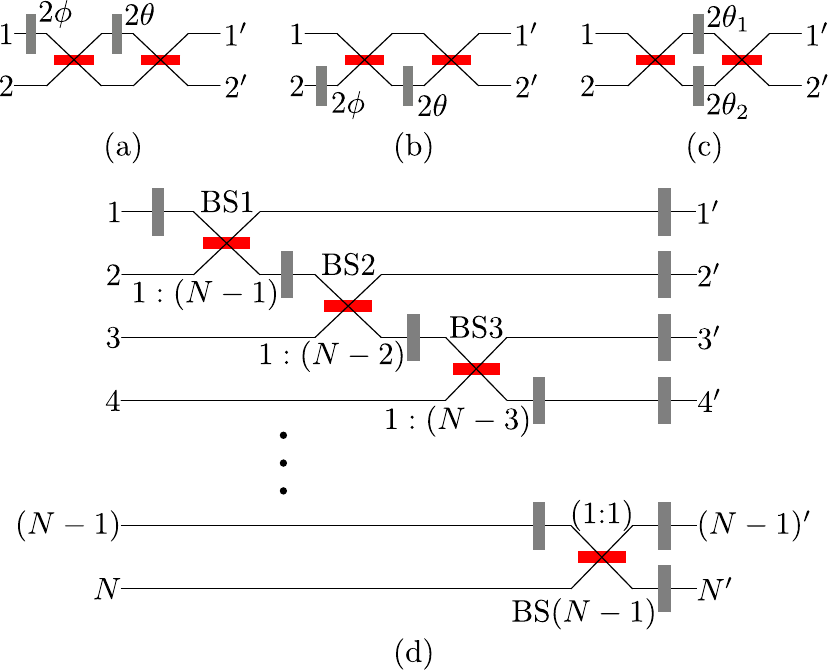}
\caption{Panels (a)-(c) depict variants of two-port Mach-Zehnder interferometers (MZIs): (a) the antisymmetric MZI (aMZI) used in the Clements scheme, (b) a modified aMZI, and (c) the symmetric MZI. Panel (d) shows an $N$-dimensional controllable multiport BS (MBS), composed of PSs (thick vertical lines) and $N-1$ BSs (thick horizontal lines), with beam-splitting ratios as indicated. When a single photon enters the first input port, it exits with equal probability across all output ports.}
\label{fig-mz}
\end{figure}

As is well known, a BS is a linear device with two input ports and two output ports. A BS with reflectance ($|R|^2$) and transmittance ($|T|^2$) ratio given by $|R|^2:|T|^2=(1-\eta):\eta$ is written as~\cite{campos89}
\begin{align} \label{bs1}
B(\eta) &= \begin{bmatrix}
\sqrt{1-\eta} & \sqrt{\eta} \\
\sqrt{\eta} & -\sqrt{1-\eta}
\end{bmatrix}.
\end{align}
Evidently, the standard 50:50 BS is denoted by $B(1/2)$. The $2 \times 2$ matrix responsible for nullifying any given off-diagonal entry is\,[see Figure\,\ref{fig-mz}\,(a)]~\cite{reck94,clements2016}
\begin{align} \label{bs1b}
e^{i\theta} 
\begin{bmatrix}
e^{i2\phi} \cos \theta & i \sin \theta \\
i e^{i2\phi} \sin \theta & \cos \theta
\end{bmatrix} &= e^{i(\theta+\phi)} \exp {(i\theta \sigma_x)} \exp {(i\phi \sigma_z)} \nonumber \\
&\equiv T(\theta,\phi).
\end{align}
Here, $\sigma_x$ and $\sigma_z$ are the standard $2 \times 2$ Pauli matrices. This was the asymmetric Mach-Zehnder interferometer\,(aMZI) considered in the Clements scheme. When both PSs are kept in the second mode instead of the first mode, the $T$-matrix assumes the form\,[see Figure\,\ref{fig-mz}\,(b)] 
\begin{align} \label{bs1c}
e^{i2(\theta+\phi)} T(-\theta,-\phi) = \tilde{T}(\theta,\phi). 
\end{align}
We shall be interested in these two types of aMZIs alone. It is also possible and potentially useful to consider a symmetric MZI (sMZI)\,[see Figure\,\ref{fig-mz}\,(c)]~\cite{bell2021}. However, we show that such sMZIs alone are not universal in Appendix\,A.
%Appendix\,\ref{mz}.

The notion of a BS can be naturally generalized to higher dimensions. For example, a tritter~\cite{zukowski97} is a three-mode linear optical device with three input and three output ports. It can be constructed using standard two-port BSs. One possible implementation involves placing a BS with transmittance $\eta=2/3$ between the first and second input modes, followed by a second BS with $\eta=1/2$ between the second and third modes. This configuration results in the following tritter transformation matrix:
\begin{align} \label{bs2}
\begin{bmatrix}
1 & 0 & 0 \\
0 & \frac{1}{\sqrt{2}} & \frac{1}{\sqrt{2}} \\
0 & \frac{1}{\sqrt{2}} & \frac{-1}{\sqrt{2}}
\end{bmatrix} 
\begin{bmatrix}
\frac{1}{\sqrt{3}} & \frac{\sqrt{2}}{\sqrt{3}} & 0 \\
\frac{\sqrt{2}}{\sqrt{3}} & \frac{-1}{\sqrt{3}} & 0 \\
0 & 0 & 1
\end{bmatrix} = \begin{bmatrix}
\frac{1}{\sqrt{3}} & \frac{\sqrt{2}}{\sqrt{3}} & 0 \\
\frac{1}{\sqrt{3}} & \frac{-1}{\sqrt{6}} & \frac{1}{\sqrt{2}} \\
\frac{1}{\sqrt{3}} & \frac{-1}{\sqrt{6}} & \frac{-1}{\sqrt{2}} \\
\end{bmatrix}.
\end{align}
It can be observed that when a unit intensity of light is injected into the first input port of the tritter, the output intensity is evenly distributed, with each port receiving $1/3$ of the total intensity. Alternatively, both BSs in the tritter can be chosen with $\eta=1/2$ to achieve a different, but still unitary, transformation. More generally, an $N$-port BS can be synthesized using only standard two-port BSs, as illustrated in Figure~\ref{fig-mz}(d)~\cite{loock2002}.

Experimentally, a tritter can be implemented in two main ways. The first relies on the standard decomposition into BSs and PSs~\cite{zukowski97,loock2002}, while the second uses a {\it single-element} device fabricated via femtosecond laser waveguide writing~\cite{spagnolo2013}. Indeed, as noted in Ref.~\cite{spagnolo2013}, the latter approach allows for a simultaneous three-photon interaction, without the need to further decompose the process into a sequence of two-mode interactions\,(BSs) and PSs. In addition, this single-element device can help implement compact and robust multiport circuits~\cite{peruzzo2011}. Therefore, the second method, in turn, greatly reduces {\it insertion loss} compared to the first~\cite{arends2024,huang2025}. 
%Furthermore, a tritter realized by femtosecond laser writing (as a single-element) will incur less insertion loss than the one realized using the standard two-mode BSs and PSs~\cite{spagnolo2013}. 

\section{Linear Optical Decompositions} \label{lo}

Now we briefly review both the Reck and the Clements schemes. We also remark on the schemes by Bell-Walmsley~\cite{bell2021} and Arends~{\it et. al.}~\cite{arends2024}. First, in the Reck and the Clements schemes, each off-diagonal entry of the target unitary is sequentially nullified using two BSs and two PSs, as illustrated in Figure \ref{fig-mz}\,(a). Since an $N$-dimensional matrix consists of $N(N-1)/2$ (lower) off-diagonal elements, we require $N(N-1)$ BSs and $N(N-1)$ PSs. Upon nullifying all off-diagonal entries, we are left with an $N$-dimensional diagonal matrix denoted by 
\begin{align} \label{lo1}
D_N(\delta_1,\ldots,\delta_N) = {\rm diag}\,\{\exp (2i\delta_1), \ldots, \exp (2i\delta_N)\}.
\end{align}
This diagonal matrix is realized using $N$ PSs. Therefore, both schemes need $N(N-1)$ BSs and $N^2$ PSs to realize any given $N$-dimensional unitary matrix. 

\begin{figure}[bp]
\centering
\includegraphics[scale=0.59]{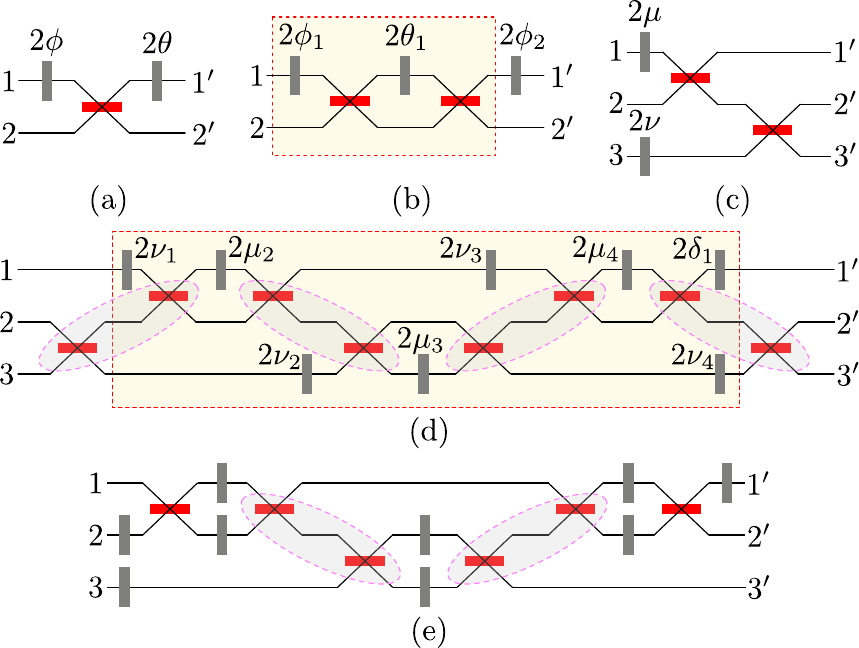}
\caption{(a) The elementary 2-dimensional block illustrating our theorem in the two-dimensional case. (b) The shaded region nullifies the off-diagonal element of a given two-dimensional unitary, while the remaining PS accounts for the diagonal matrix. (c) Elementary $3 \times 3$ tritter block composed of two PSs and two BSs, each with reflectivity $\eta=1/2$. (d) Proposed scheme for realizing an arbitrary $U_3$ using four fixed tritter blocks (shown in rotated ellipses). The shaded rectangular region implements any desired $U_3$\,[see Eq.\,(\ref{td1b})], following the Clements decomposition scheme. (e) Bell-Walmsley-Clements scheme that uses two fixed tritters and two BSs to realize any $U_3$.}
\label{fig-tri}
\end{figure}

In the Reck scheme, the last row of the unitary [except the $(N,N)$-th entry] is first nullified. Then the $(N-1)$-th row [except the $(N-1,N-1)$-th entry] is completely nullified. Proceeding further, one can obtain the $N$-dimensional diagonal matrix that can be realized using $N$ PSs. This nullification procedure gives rise to a triangular architecture. In contrast, nullification of the unitary in the Clements scheme occurs in the following order: $(N,1)$-th term, $(N-1,1)$-th term, $(N,2)$-th term, $(N,3)$-th term, $(N-1,2)$-th term, and so on. This form of nullification leads to a rectangular architecture, which reduces the optical depth of the multiport interferometer and thereby enhances its tolerance to optical losses. Therefore, any $N$-dimensional unitary $U_N$ can be compactly written as
\begin{align} \label{lo2}
U_N = D_N \prod_{m,n}^{R/C} {\rm T}_{mn}(\theta_{mn},\phi_{mn}).     
\end{align}
Here, ${\rm T}_{mn}(\theta_{mn},\phi_{mn})$ is an $N$-dimensional identity matrix with matrix entries at positions $(m,m)$, $(m,n)$, $(n,m)$ and $(n,n)$ being replaced by the entries of $T(\theta_{mn},\phi_{mn})$ in Eq.\,(\ref{bs1b})~\cite{reck94}. Also, the specific order in which different ${\rm T}_{mn}$-matrices are to be multiplied is dictated by Reck\,($R$) or Clements\,($C$) schemes. We remark that we can choose any one of the phases in $D_N$ to be zero without loss of generality, as we cannot measure the overall phase. For instance, the Clements scheme for the 3-dimensional case is
\begin{align} \label{lo3}
\!\!\! U_3 = D_3(\delta_1,0,\delta_3) {\rm T}_{12}(\theta_3,\phi_3) {\rm T}_{23}(\theta_{2},\phi_{2})  {\rm T}_{12}(\theta_1,\phi_1).   
\end{align}
Bell and Walmsley~\cite{bell2021} introduced an alternative realization of the Reck and Clements interferometers, where the basic components are sMZIs combined with external PSs. In Appendix~\ref{mz}, we discuss the role of external PSs in ensuring the universality of this scheme.
% and illustrate the Bell-Walmsley-Reck interferometer for $N=3$

Arends~{\it et al.}\cite{arends2024} proposed a new decomposition scheme motivated by the emerging feasibility of fabricating integrated devices that implement $m$-dimensional unitaries. This approach allows a given $N$-dimensional unitary to be constructed from smaller subunits of dimensions $m_1 \times m_1$, $m_2 \times m_2$, and so on, where each $m_i<N$ and $m_i \geq 2$. In contrast, the Clements scheme builds a $U_N$ using only two BSs and two PSs per unit cell [see Figure~\ref{fig-mz}(a)]. Their numerical simulations show that the use of larger building blocks ($m>2$) can enhance the fidelity, highlighting the advantage of incorporating higher-dimensional components.

We stress that none of the linear optical decompositions realizing a generic $U_N$ can surpass the lower bound set forth by the Reck scheme in terms of the number of BSs and PSs. This can be attributed to the fact that the Reck scheme needs exactly $N^2$ PSs to realize any arbitrary $N$-dimensional unitary represented by $N^2$ independent real parameters~\cite{bala2010}. Therefore, the Reck scheme is optimal in this sense. 

\section{Finite Identical-Multiport Decompositions} \label{fi}

In this Section, we show that any $N$-dimensional unitary can be realized as a finite sequence of MBSs and PSs. We begin by presenting the explicit constructions for the two- and three-dimensional cases. For $N \geq 4$, the decomposition follows from the BWC framework, after which we briefly outline its connection to current experimental implementations. The construction relies on the following theorem from the theory of Lie groups~\cite{pontrjagin46}.

{\noindent \bf Theorem:} Any element in a connected group can be represented as a finite product of elements belonging to the group.

A topological proof of this theorem, along with its relevance to the MBS framework, is provided in Appendix\,\ref{tog}. We now illustrate the theorem for the case $N=2$. The basic building block, consisting of one BS and two PSs, is shown in Figure~\ref{fig-tri}(a). By cascading two such blocks as depicted in Figure~\ref{fig-tri}(b), we can realize an arbitrary $U_2$. Specifically, the linear optical elements within the shaded region---corresponding to the transformation $T(\theta_1,\phi_1)$ [see Eq.~(\ref{bs1b})]---serve to nullify the off-diagonal element of the target $U_2$. Since the resulting matrix must still be unitary and the global phase is physically irrelevant, the remaining transformation is a diagonal unitary $D_2(\phi_2,0)$. We have
\begin{align} \label{su2}
D_2(\phi_2,0) T(\theta_1,\phi_1) &= \exp {[i(\theta_1+\phi_1+\phi_2)]} \exp {(i\phi_2 \sigma_z)} \nonumber \\
&\,\,\,\times \exp {(i\theta_1 \sigma_x)} \exp {(i\phi_1 \sigma_z)}.
\end{align}
Discarding the overall phase factor, we find that the right-hand side spans the entire SU(2).
%This remaining $2 \times 2$ diagonal matrix is realized using the PS outside the shaded region. 
Hence, the theorem holds for $N=2$. 

\begin{figure}[tbp]
\centering
\includegraphics[scale=0.5]{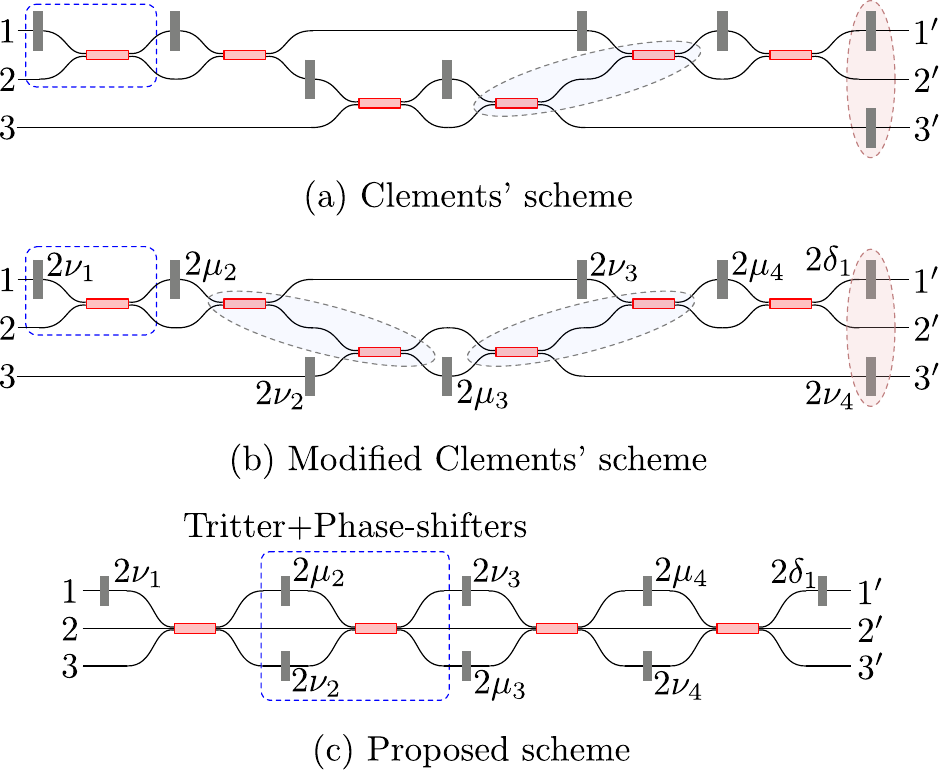}
\caption{We assume that each BS or tritter can be fabricated as a single integrated block. Under this assumption, the realization of any $U_3$ requires: (a) in the original Clements scheme, either 6 identical blocks or 5 blocks with optimized placement; (b) in the modified Clements scheme, 4 non-identical blocks with varied configurations; (c) in the proposed scheme shown in Figure~\ref{fig-tri}(d), only 4 {\it identical} fixed tritter blocks. In both (a) and (b), tritters are shown in rotated ellipses.}
\label{fig-ps}
\end{figure}

In the 3-dimensional case, the elementary building block is a tritter\,[see Figure\,\ref{fig-tri}\,(c)], with both BSs being $B(1/2)$. The Lie group theorem ensures that we can realize any given $U_3$ using a sequence of such tritter blocks. In Fig.\,\ref{fig-tri}\,(d), it is shown that 4 tritters can realize any given $U_3$. Because of the Clements scheme, we know that any given $U_3$ can be realized using the linear optical elements inside the rectangular shaded region. Now suppose that we want to realize any $U_3$. Then we have to find out the corresponding Clements decomposition for  
\begin{align} \label{td1}
U_{\rm BS} U_3 U_{\rm BS} \equiv U'_3,    
\end{align}
where $U_{\rm BS}$ is the BS matrix mixing modes 2 and 3. Note that we have set $\mu_1=0$. Unlike Eq.\,(\ref{lo3}), the modes 2 and 3 are connected by the aMZI shown in Figure\,\ref{fig-mz}\,(b). So, the decomposition corresponding to Figure\,\ref{fig-tri}\,(d) is
\begin{align} \label{td1b}
\!\!\!\! U_3 = D_3(\delta_1,0,\nu_4) {\rm T}_{12}(\mu_4,\nu_3) \tilde{{\rm T}}_{23}(\mu_{3},\nu_{2})  {\rm T}_{12}(\mu_2,\nu_1),  
\end{align}
where $\tilde{{\rm T}}_{23}$ is a 3-dimensional identity matrix whose entries at positions (2,2), (2,3), (3,2), and (3,3) being replaced by $\tilde{T}$ in Eq.\,(\ref{bs1c}). Thus, our decomposition requires 4 {\it identical} fixed tritter blocks, or 2 elementary blocks made of 2 fixed tritters. We note that the BWC scheme, for $N=3$, employs two fixed tritters and two BSs\,[see Figure\,\ref{fig-tri}\,(e)]. %{Although both schemes require five phase masks, the latter requires nine PSs.}

In contrast, the Clements scheme for a $U_3$ employs 6 BS blocks\,[see Figure\,\ref{fig-ps}\,(a)]. It can be observed that the two BSs can be combined to form a single element, thereby potentially reducing the insertion loss~\cite{huang2025,spagnolo2013}~\footnote{As an analogy, in gate-based quantum computation, a Toffoli gate~\cite{nielsen2010} can either be implemented directly in a single step or decomposed into a sequence of Hadamard, phase, and controlled-NOT gates. However, because each gate introduces imperfections and errors, the latter approach accumulates more noise. This highlights the need for more compact quantum circuits.} We have demonstrated this decomposition for the $U_3$ arising in the case of USD of two non-orthogonal qubit states in Appendix\,\ref{po}. In Appendix\,\ref{op}, the same USD unitary is employed to show the optimality of our decomposition, i.e., three tritters are not sufficient to realize any 3-dimensional unitary. 

\begin{figure}[htbp]
\centering
\includegraphics[scale=0.38]{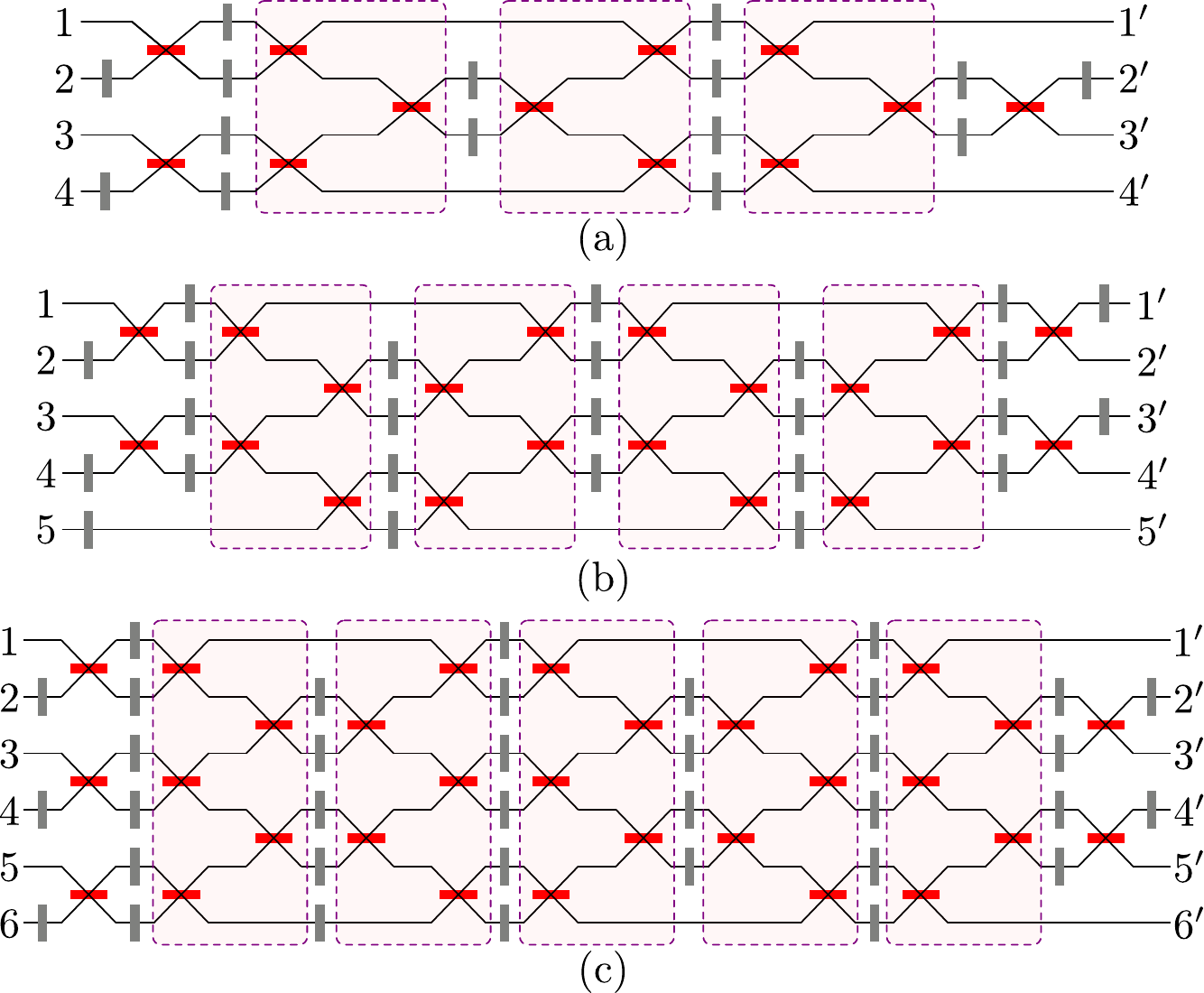}
\caption{Panels (a)-(c) illustrate the Bell-Walmsley-Clements (BWC) decomposition for dimensions $N=4, 5$, and $6$ respectively. Each shaded region represents an MBS corresponding to that dimension. This architecture clearly demonstrates the $\mathcal{O}(N)$ linear scaling of fixed identical MBS blocks required for universal unitary realization, needing $N+2$ phase masks in total.}
\label{fig-bw}
\end{figure}
%Panels (a)--(c) illustrate the Bell-Walmsley-Clements (BWC) decomposition for dimensions $N=4$, 5, and 6, respectively. Each shaded region represents a MBS corresponding to that dimension. In this framework, any unitary matrix can be realized as a sequence of fixed MBSs and BSs, requiring $N+2$ phase masks in total.

For $N \geq 4$, the solution is remarkably simple. The BWC framework indicates that any unitary $U_N$ can be realized using identical MBSs, as illustrated in Fig.~\ref{fig-bw}. It follows that the implementation requires $N+2$ phase masks, $N-1$ fixed MBSs, and $N-1$ BSs. While Refs.~\cite{saygin2020,pereira2025} conjectured that $N+2$ phase masks are necessary for multichannel block-based decompositions employing a DFT matrix as a single block, we demonstrate that the same number of phase masks suffices when using identical fixed MBSs and BSs. In addition, we have shown that $\mathcal{O}(N)$ fixed identical MBSs and BSs are sufficient to realize any $U_N$, without increasing the number of tunable PSs required by the standard Clements scheme, which requires $\mathcal{O}(N^2)$ fixed BSs.

We now highlight a few connections with the ongoing research and outline some potential applications of our results. In Ref.~\cite{huang2025}, multiphoton quantum interference of a topology-optimized tritter has been characterized through single- and two-photon statistics. Recently, a $4 \times 4$ MBS has been utilized for a boosted Bell-state measurement scheme~\cite{hauser2025}. Moreover, multimode couplers of dimensions $2 \times 2$ and $4 \times 4$ offer substantial simplification and concatenation of photonic quantum circuits~\cite{peruzzo2011}. Finally, three-photon interference in an ultrafast laser-written tritter was experimentally observed~\cite{spagnolo2013}. Building on these advances, we expect our proposed decomposition to enable robust and compact multiport linear-optical architectures~\cite{peruzzo2011}. By employing MBSs instead of cascaded two-mode BSs and PSs, our approach reduces insertion loss and thereby enhances the overall performance of multiport interferometers~\cite{spagnolo2013, huang2025,arends2024}.

\section{Concluding Remarks} \label{cr}

We have proposed a reconfigurable linear optical setup to realize any given unitary matrix in a given dimension. The universality of this decomposition is grounded in the Lie group theorem. While the case of $N=2$ is straightforward, proving universality for $N=3$ using four tritters requires a bit of effort. Saygin~\textit{et al.}\cite{saygin2020} were unable to demonstrate universality for $N=3$ with static multi-channel blocks; in contrast, we show that four {\it identical} fixed tritters\,(or multi-channel blocks~\cite{saygin2020} or multimode blocks~\cite{arends2024}) suffice. Furthermore, we have shown that within the BWC framework, $N+2$ phase masks, $N-1$ fixed MBSs, and $N-1$ BSs are sufficient to realize any unitary $U_N$ for $N \geq 3$. In particular, our scheme merely requires $\mathcal{O}(N)$ fixed {\it identical} MBSs and $\mathcal{O}(N)$ fixed BSs for an $N$-dimensional unitary realization, whereas the Clements scheme requires $\mathcal{O}(N^2)$ fixed BSs. This substantial reduction in fixed-component complexity is achieved while maintaining the same number of tunable PSs. Crucially, the uniformity of the MBS components offers a distinct advantage for scalable monolithic integration. Finally in Appendix\,\ref{op}, we present a reconfigurable setup implementing the unitary arising from the USD problem for two nonorthogonal qubit states. 

Assuming that we can mass-fabricate MBSs in the near future, we believe that our decomposition will have an advantage over the existing reconfigurable linear optical schemes in terms of loss tolerance. Moreover, our approach minimizes insertion losses and mitigates the impact of fabrication imperfections by enabling the linear-optics structure to be laser-written onto optical chips as a monolithic element. We have established the theorem for $N \geq 4$ within the BWC framework. Other choices of fixed or controllable MBSs can also be employed, as the theorem guarantees that any finite set of identical MBS elements suffices for the decomposition (see Appendix~\ref{tog}). However, the optical path length of the proposed scheme may exceed that of the Clements architecture~\cite{clements2016}, and this effect warrants further investigation. 
%We have demonstrated the theorem for $N \geq 4$ within the BWC framework. One can also consider other fixed or controllable MBSs, as the theorem ensures us that any finite number of identical MBS elements will be sufficient for the decomposition\,(see Appendix\,\ref{tog}). Nonetheless, the optical path length~\cite{clements2016} of the proposed scheme might exceed the same of the Clements scheme. This should be explored further.
%One has to analyze the higher-dimensional case in detail in terms of error metrics and by introducing absorption in the BSs. When we introduce MBSs for $N>3$, 

%%%%%%%%%%%%%%%%%%%%%%%%%%%%%%%%%%%%%%%%%%%%%%%%%%%%%%%%%%%%%%%%%%%%%%%%%%%%%%%%%%%%%%%%%%%%%%%%%%%%%%%%%%%%%%%%%%%%%%%%%%%%%%%%%%%%%%%%%%%%%%%%%%%%%%%%%%%%%%%%%%%%%%%%%%%%%%%%%%%%

\begin{acknowledgments}
We acknowledge support from the Federal Ministry of Education and Research in Germany (BMBF, project PhotonQ: FKZ 13N15758). We thank Pradip Laha, Jeldrik Huster, Vincent Girouard, and Nicolas Quesada for their insightful comments and discussions. We specifically thank Daniel Bhatti for helpful discussions on the universality proof.
\end{acknowledgments}

\appendix

\section{(Non-)universality of sMZI} \label{mz}
% and the Bell-Walmsley scheme

A symmetric MZI can be described by the following $2 \times 2$ matrix
\begin{align} \label{mz1}
B(1/2) M_2 B(1/2) &= \frac{1}{2}
\begin{bmatrix}
e^{i\phi_{11}} + e^{i\phi_{12}} & e^{i\phi_{11}} - e^{i\phi_{12}} \\
e^{i\phi_{11}} - e^{i\phi_{12}} & e^{i\phi_{11}} + e^{i\phi_{12}}
\end{bmatrix} \nonumber \\
&= \exp {\left[ \frac{i}{2} (\phi_{11}+\phi_{12}) \right]} \nonumber \\
&\,\,\,\times \exp { \left[ \frac{i}{2} (\phi_{11}-\phi_{12}) \sigma_x \right]}.
\end{align}
Because this contains the $\sigma_x$-term alone, we cannot realize any general element in SU(2), regardless of how many concatenated sMZIs are kept. In contrast, an aMZI along with a PS can realize any SU(2) element\,[see Figure\,2\,(b) and Eq.\,(8) in the main file]. We now formally demonstrate the non-universality of sMZIs by presenting a counterexample. Specifically, we show that the aMZI blocks used in the Reck or Clements schemes cannot be replaced with sMZI blocks. As an illustrative case, consider the following 4-dimensional unitary:
\begin{align} \label{mz2}
\begin{bmatrix}
U_2 & 0 \\
0 & \exp{(-i\theta \sigma_y)}
\end{bmatrix}.   
\end{align}
Here, $\sigma_y$ is the Pauli matrix. The (4,3)-th element can be nullified when
\begin{align} \label{mz3}
\sin \theta (e^{i\phi_{11}} + e^{i\phi_{12}}) + \cos \theta (e^{i\phi_{11}} - e^{i\phi_{12}}) = 0 \nonumber \\
\,\,\,{\rm (or)}\,\,\,
e^{i(\phi_{11}-\phi_{12})} = \frac{\cos \theta - \sin \theta}{\cos \theta + \sin \theta}.
\end{align}
Because the RHS is real for any $\theta$, no solution exists. Therefore, we cannot nullify the (4,3)-th element of this unitary using the sMZI block in Eq.\,(\ref{mz1}). We remark that an sMZI in two modes is equivalent to keeping just one PS in one of the modes, as the overall phase cannot be measured experimentally. While a network of sMZI interferometers alone is not universal, universality can be restored by incorporating external PSs~\cite{bell2021}.

\section{USD problem and the 3-dimensional unitary} \label{po}

We now turn our attention to a specific application where these linear optical decompositions show significant promise: unambiguous state discrimination (USD). It has been shown that positive operator-valued measure (POVM) schemes can achieve higher success probabilities for distinguishing two non-orthogonal quantum states than projective measurements~\cite{ivanovic87,peres88,chefles2010}. In the original proposal, the required unitary acting on the system and ancilla was four-dimensional. Subsequent work demonstrated that the same optimal success probability could be achieved using only a three-dimensional unitary~\cite{loock2006}. In this scenario, two of the three POVM outcomes are conclusive, while the third yields an inconclusive result. Now, we present an explicit reconfigurable linear optical setup that realizes this three-dimensional USD unitary.

We consider two nonorthogonal qubit states represented by
\begin{align}
|\chi_\pm \rangle = a|\bar{0} \rangle \pm b|\bar{1} \rangle + 0 |\bar{2} \rangle,    
\end{align}
where $a$ and $b$ are two real numbers with $a>b$ and $a^2+b^2=1$. Also, $\{|\bar{0} \rangle, |\bar{1} \rangle, |\bar{2} \rangle\}$ are orthogonal basis states, and the state $|\bar{2} \rangle$ has been included for later convenience. The question is how to distinguish the two given states $|\chi_\pm \rangle$ using the POVM measurements. In the POVM formalism, we require $\rho_A \otimes \rho_B \in \mathcal{H}_A \times \mathcal{H}_B$. But we can also use the approach $\rho_A \oplus \rho_B \in \mathcal{K} \oplus \mathcal{K}^\perp$, where $\mathcal{K}$ is the space belonging to $\rho_A$ and $\mathcal{K}^\perp$ is a one-dimensional subspace orthogonal to $\mathcal{K}$. In the former one, there is too much `redundancy' in the Hilbert space, whereas the latter one is rather `tight'. In order to perform the optimal USD, it was shown that~\cite{loock2006} the following POVM operators would be required\,:
\begin{align}
\hat{E}_\mu &= |u_\mu \rangle \langle u_\mu|, \,\,\,\mu=1,2,3, \label{emu} \\
{\rm where}\,\,\,
|u_{1/2} \rangle &= \frac{1}{\sqrt{2}} \frac{b}{a} |\bar{0} \rangle \pm \frac{1}{\sqrt{2}} |\bar{1} \rangle, \label{u12} \\
{\rm and} \,\,\,
|u_3 \rangle &= \sqrt{1-\frac{b^2}{a^2}} |\bar{0} \rangle. \label{u3}
\end{align}
While both POVM operators, $\hat{E}_1$ and $\hat{E}_2$, give unambiguous outcomes, namely, 
\begin{align*}
{\rm Tr}\,(\hat{E}_1 |\chi_- \rangle \langle \chi_-|)=0 \,\,\,{\rm and}\,\,\, {\rm Tr}\,(\hat{E}_2 |\chi_+ \rangle \langle \chi_+|)=0,
\end{align*} 
the outcome due to the third POVM operator is inconclusive. The corresponding states in the extended Hilbert space are
\begin{align}
|w_\mu \rangle &= |u_\mu \rangle + |N_\mu \rangle, \label{wmu} \\
{\rm where}\,\,\,
|N_{1/2} \rangle &= \frac{1}{\sqrt{2}} \sqrt{1-\frac{b^2}{a^2}} |\bar{2} \rangle, \label{n12} \\
{\rm and}\,\,\,
|N_3 \rangle &= -\frac{b}{a} |\bar{2} \rangle. \label{n3}
\end{align}

Now it is easy to write down the unitary\,($U_{\rm USD}$) in the extended Hilbert space as
\begin{align} \label{po1a}
U_{\rm USD} = \sum_{j=0}^2 |\bar{j} \rangle \langle w_j| = 
\begin{bmatrix}
\frac{1}{\sqrt{2}} \delta & \frac{1}{\sqrt{2}} & \frac{1}{\sqrt{2}} \sqrt{1-\delta^2} \\
\frac{1}{\sqrt{2}} \delta & \frac{-1}{\sqrt{2}} & \frac{1}{\sqrt{2}} \sqrt{1-\delta^2} \\
\sqrt{1-\delta^2} & 0 & -\delta
\end{bmatrix},
\end{align}
where $\delta=b/a$ with $a,b \in \mathbb{R}$. When this unitary acts on the state $|\chi_\pm \rangle$, we obtain
\begin{align} \label{po2a}
\begin{bmatrix}
\frac{1}{\sqrt{2}} \delta & \frac{1}{\sqrt{2}} & \frac{1}{\sqrt{2}} \sqrt{1-\delta^2} \\
\frac{1}{\sqrt{2}} \delta & \frac{-1}{\sqrt{2}} & \frac{1}{\sqrt{2}} \sqrt{1-\delta^2} \\
\sqrt{1-\delta^2} & 0 & -\delta
\end{bmatrix} 
\begin{bmatrix}
a \\
\pm b \\
0
\end{bmatrix} = 
\begin{bmatrix}
\frac{\delta a \pm b}{\sqrt{2}} \\
\frac{\delta a \mp b}{\sqrt{2}} \\
a \sqrt{1-\delta^2}
\end{bmatrix}. 
\end{align}
With this, the output state is either $[\sqrt{2} b , 0, \sqrt{a^2-b^2}]^\top$ or $[0, \sqrt{2} b , \sqrt{a^2-b^2}]^\top$.
Let us introduce the following notation: $P(|i \rangle|j)$ is the probability of measuring the state ``$|i \rangle$'' given an outcome ``$j$''. Then 
%$P(|0 \rangle|+)$ is
\begin{align} 
P(|0 \rangle|+) &= |\langle 0|U|\chi_+ \rangle|^2 = 2b^2, \label{po2b} \\
{\rm and}\,\,\,
P(|1 \rangle|-) &= |\langle 1|U|\chi_- \rangle|^2 = 2b^2. \label{po2c}
\end{align}
Now the total success probability is
\begin{align} \label{po3b}
P_{\rm success} = \frac{1}{2} \times P(|0 \rangle|+) + \frac{1}{2} \times P(|1 \rangle|-) = 2b^2.
\end{align}

Suppose we send in the states $|\chi_\pm \rangle$ through the 3-dimensional linear optical setup realizing $U_{\rm USD}$\,(multiple-rail encoding). Then the probability of unambiguously distinguishing the qubit states is $2b^2$, which coincides with the optimal, maximal success probability of $1- \langle \chi_+|\chi_- \rangle=1-(a^2-b^2)=2b^2$. For the multiple-rail encoding, we identify that $|\bar{0} \rangle=|100 \rangle$, $|\bar{1} \rangle=|010 \rangle$, and $|\bar{2} \rangle=|001 \rangle$. Here, for instance, $|100 \rangle$ denotes that a single photon is sent through the first input port alone, not the remaining two ports. The Clements decomposition realizing $U_{\rm USD}$ given in Eq.\,(\ref{po2a}) is 
\begin{align} \label{po4a}
U_{\rm USD} &= -e^{i\pi/4}
\begin{bmatrix}
\frac{e^{-i\pi/4}}{\sqrt{2}} & \frac{e^{i\pi/4}}{\sqrt{2}} & 0 \\
\frac{-e^{-i\pi/4}}{\sqrt{2}} & \frac{e^{i\pi/4}}{\sqrt{2}} & 0 \\
0 & 0 & 1
\end{bmatrix} \nonumber \\
&\,\,\,\times
\begin{bmatrix}
1 & 0 & 0 \\
0 & e^{-i(\theta_3+3\pi/4)} \delta & -ie^{-i(\theta_3+3\pi/4)} \sqrt{1-\delta^2} \\
0 & -i e^{-i\theta_3} \sqrt{1-\delta^2} & e^{-i\theta_3} \delta
\end{bmatrix} \nonumber \\
&\,\,\,\times
\begin{bmatrix}
1 & 0 & 0 \\
0 & 1 & 0 \\
0 & 0 & e^{i(\theta_3-\pi/4)}  
\end{bmatrix} 
\begin{bmatrix}
0 & -1 & 0 \\
-e^{i(\theta_3+\pi/4)} & 0 & 0 \\
0 & 0 & 1
\end{bmatrix},  
\end{align}
where $\theta_3=\tan^{-1} (\sqrt{1-\delta^2}/\delta)$. Making use of the identity
\begin{align} \label{po4b}
T^{-1}(\theta,\phi) D_2(\alpha,0) = D_2(-\phi,0) T(-\theta,\alpha),    
\end{align}
Eq.\,(\ref{po4a}) can be rewritten as
\begin{align} \label{dc11a}
U_{\rm USD} &= e^{i(\theta_3+\pi/4)} D_3(-\pi/4,0, 3\pi/8) \nonumber \\
&\,\,\,\times {\rm T}_{12}(-3\pi/4,\pi/2-\theta_3/2) {\rm T}_{23}(-\theta_{3},\pi/8-\theta_3/2) \nonumber \\
&\,\,\,\times {\rm T}_{12}(\pi/2,\theta_3/2+\pi/8).    
\end{align}
% \begin{align} \label{dc11a}
% U_{\rm USD} &= e^{i(\theta_3+\pi/4)} \times {\rm diag}\,\{e^{-i\pi/2},1, e^{i3\pi/4} \} \nonumber \\
% &\,\,\,\times {\rm T}_{2}(-3\pi/4,\pi-\theta_3) {\rm T}_{3}(-\theta_{3},\pi/4-\theta_3) \nonumber \\
% &\,\,\,\times {\rm T}_{1}(\pi/2,\theta_3+\pi/4).    
% \end{align}
Here, $e^{i(\theta_3+ \pi/4)}$ is an overall phase factor and can be safely ignored. Finally, in order to find out the tritter decomposition of $U_{\rm USD}$, we let $U_3=U_{\rm USD}$ in Eq.\,(\ref{td1}) of the main file to obtain
\begin{widetext}
\begin{align} \label{td2}
U'_3 &= \begin{bmatrix}
\frac{1}{\sqrt{2}} \delta & \frac{1}{2}(1 + \sqrt{1-\delta^2}) & \frac{1}{2}(1 - \sqrt{1-\delta^2}) \\
\frac{\delta}{2} + \frac{1}{\sqrt{2}} \sqrt{1-\delta^2} & -\frac{1}{2\sqrt{2}} + \frac{1}{2\sqrt{2}} \sqrt{1-\delta^2} - \frac{1}{2} \delta & -\frac{1}{2\sqrt{2}} - \frac{1}{2\sqrt{2}} \sqrt{1-\delta^2} + \frac{1}{2} \delta \\
\frac{\delta}{2} - \frac{1}{\sqrt{2}} \sqrt{1-\delta^2} & -\frac{1}{2\sqrt{2}} + \frac{1}{2\sqrt{2}} \sqrt{1-\delta^2} + \frac{1}{2} \delta & -\frac{1}{2\sqrt{2}} - \frac{1}{2\sqrt{2}} \sqrt{1-\delta^2} - \frac{1}{2} \delta
\end{bmatrix}.
\end{align}
\end{widetext}
The Clements decomposition corresponding to $U'_3$ is
\begin{align} \label{td3}
U'_3 &= \begin{bmatrix}
-i e^{-i\theta_2} \frac{(1+\sqrt{1-\delta^2}-\sqrt{2} \delta)}{A_1} & -e^{-i\theta_2} \frac{\sqrt{2} (1-\sqrt{1-\delta^2})}{A_1} & 0 \\
-i e^{-i\theta_2} \frac{\sqrt{2} (1-\sqrt{1-\delta^2})}{A_1} & e^{-i\theta_2} \frac{(1+\sqrt{1-\delta^2}-\sqrt{2} \delta)}{A_1} & 0 \\
0 & 0 & 1
\end{bmatrix} \nonumber \\
&\,\,\,\times \begin{bmatrix}
1 & 0 & 0 \\
0 & e^{-i\theta_3} \frac{(1+\sqrt{2}\delta+\sqrt{1-\delta^2})}{2\sqrt{2}} & ie^{-i\theta_3} \frac{A_1}{2\sqrt{2}} \\
0 & -e^{-i(\theta_2+\theta_3)} \frac{A_1}{2\sqrt{2}} & i e^{-i(\theta_2+\theta_3)} \frac{(1+\sqrt{2}\delta+\sqrt{1-\delta^2})}{2\sqrt{2}}
\end{bmatrix} \nonumber \\
&\,\,\,\times 
\begin{bmatrix}
e^{i(\theta_2-\theta_1+\pi)} & 0 & 0 \\
0 & e^{i(\pi+\theta_3+\theta_2-\theta_1)} & 0 \\
0 & 0 & e^{i(\pi/2+\theta_2+\theta_3)}
\end{bmatrix} \nonumber \\
&\,\,\,\times
\begin{bmatrix}
-i e^{i\theta_1} \frac{(\sqrt{1-\delta^2}-1+\sqrt{2}\delta)}{A_1} & i e^{i\theta_1} \frac{\sqrt{2}(\delta-\sqrt{2}\sqrt{1-\delta^2})}{A_1} & 0 \\
e^{i\theta_1} \frac{\sqrt{2}(\delta-\sqrt{2}\sqrt{1-\delta^2})}{A_1} & e^{i\theta_1} \frac{(\sqrt{1-\delta^2}-1+\sqrt{2}\delta)}{A_1} & 0 \\
0 & 0 & 1
\end{bmatrix}.     
\end{align}
Here, $\phi_1=-\pi/2$, $\theta_1= \tan^{-1} \left( \frac{\sqrt{2}\delta -2\sqrt{1-\delta^2}}{\sqrt{1-\delta^2}-1+\sqrt{2} \delta} \right)$, $\phi_2=\pi/2$, $\theta_2 = \tan^{-1} \left[ \frac{\sqrt{2}(1- \sqrt{1-\delta^2})}{1+\sqrt{1-\delta^2} -\sqrt{2} \delta} \right]$, $\phi_3=-\pi/2+\theta_2$, $\theta_3 = \tan^{-1} \left[ \frac{A_1}{1+\sqrt{2}\delta+\sqrt{1-\delta^2}} \right]$, and
\begin{align} \label{td4}
A_1^2 &= 6-\delta^2 -2 \sqrt{1-\delta^2} -2\sqrt{2}\delta - 2\sqrt{2} \delta \sqrt{1-\delta^2}.      
\end{align}
Now once again making use of Eq.\,(\ref{po4b}) in Eq.\,(\ref{td3}), we arrive at
\begin{align} \label{td5}
U'_3 &= e^{i(2\theta_2+\theta_3)} D_3(-\pi/4,0,\pi/4-\theta_2/2) \nonumber \\
&\,\,\, \times {\rm T}_{12}(-\theta_2, (\pi-\theta_1-\theta_2-\theta_3)/2) \nonumber \\
&\,\,\,\times 
\begin{bmatrix}
1 & 0_{1 \times 2} \\
0_{2 \times 1} & e^{-i(2\theta_3+\theta_2-\pi/2)} T(\theta_{3},\pi/4-\theta_1/2)
\end{bmatrix} \nonumber \\
&\,\,\,\times {\rm T}_{12}(\theta_1,-\pi/4).    
\end{align}
Here, $0_{j \times k}$ is a zero matrix with $j$ rows and $k$ columns.

\section{Optimality of the tritter decomposition} \label{op}

Here, we demonstrate that the USD unitary defined in Eq.\,(\ref{po1a}) cannot be realized using only three tritters. While a subset of 3-dimensional unitaries can indeed be implemented with three tritter blocks, the specific structure of the USD unitary necessitates four. To illustrate this, consider the tritter-based decomposition shown in Figure\,2\,(d) of the main file, and focus on the first three tritters from the left. We have:
\begin{align} \label{ot1a}
B_{12} D_3(\nu_3,0,0) \tilde{{\rm T}}_{23}(\mu_3,\nu_2) {\rm T}_{12}(\mu_2,\nu_1) B_{23}, 
\end{align}
where $B_{ij}$ is a 50:50 BS connecting modes $i$ and $j$. If it is possible to realize $U_{\rm USD}$ in Eq.\,(\ref{po1a}) using 3 tritters, then
\begin{align} \label{ot1b}
U_{\rm USD} &= B_{12} D_3(\nu_3,0,0) \tilde{{\rm T}}_{23}(\mu_3,\nu_2) {\rm T}_{12}(\mu_2,\nu_1)  B_{23} \nonumber \\
B_{12} U_{\rm USD} &= D_3(\nu_3,0,0) \tilde{{\rm T}}_{23}(\mu_3,\nu_2) {\rm T}_{12}(\mu_2,\nu_1) B_{23}.
\end{align}
After straightforward matrix multiplication, we can compare (1,2) and (1,3) terms on both sides and obtain
\begin{align} 
0 &= \frac{i}{\sqrt{2}} e^{i(\mu_2+\nu_3)} \sin \mu_2, \label{ot4a} \\
{\rm and}\,\,\,
\sqrt{1-\delta^2} &= \frac{i}{\sqrt{2}} e^{i(\mu_2+\nu_3)} \sin \mu_2, \label{ot4b}
\end{align}
respectively. These two equations cannot be satisfied simultaneously for a given $\delta$. Therefore, the USD unitary will require 4 tritters.

\section{Proof of the theorem} \label{tog}

Here, we provide a topological proof of the Lie group theorem. The definitions and arguments that follow are based on Refs.\,\cite{pontrjagin46, sagle73,montgomery2018}. Theorem\,15 on page no. 76 of Ref.~\cite{pontrjagin46} reads: ``{\it A connected topological group $G$ is generated by an arbitrary neighborhood $U$ of the identity. This means that $G$ coincides with the sum of all sets of the form $U^n$, $n=1,2,\cdots$, or, what is the same, that every element of $G$ can be represented as a finite product of elements belonging to $U$}''. We first present the necessary definitions and preliminary lemmas, then give two alternative proofs of the main theorem. We conclude with intuitive explanations and a detailed illustration of the theorem's relevance to the MBS element belonging to the connected topological group ${\rm SU}(N)$.

%We begin with some definitions and two preliminary lemmas, followed by two proofs of the main theorem. Next, we provide some intuitive explanations and show explicitly how this theorem can be utilized for the MBS context.

%Any member of the connected group can be generated from an arbitrary neighborhood element of the identity. In other words, the special unitary group ${\rm SU}(N)$ coincides with the sum of all sets of the form $U^n$, $n=1,2,\ldots$, where $U$ is the unitary representing the MBS. 

{\noindent \bf Definition:} A subset $X \subset S$ is called closed if the complement $S \setminus X$ is open.

{\noindent \bf Definition:} A topological space is connected if it cannot be expressed as the union of two nonempty, disjoint open subsets.

{\noindent \bf Lemma\,1:} Consider a connected group $G$. If $U \subseteq G$, then the set
\begin{align} \label{tog2a}
H \equiv \{ u_1^{\varepsilon_1} u_2^{\varepsilon_2} \cdots u_k^{\varepsilon_k}\,\,:\,\, k \geq 0, \,\,\, u_i \in U, \,\,\, \varepsilon_i \in \{-1,+1\} \}
\end{align} 
is the subgroup generated by $U$; i.e. $H$ is a subgroup.
%$U \subset H$, and any subgroup $K \leq G$ with $U \subset K$ must contain $H$.

{\noindent \bf Proof:} If $u_1^{\varepsilon_1} u_2^{\varepsilon_2} \cdots u_k^{\varepsilon_k}$ and $v_1^{\delta_1} v_2^{\delta_2} \cdots v_l^{\delta_l}$ are two elements of $H$, then the product is evidently an element of $H$ too. So, closure axiom is satisfied. Associativity is trivial. Identity belongs to $k=0$ and is an element of $H$. Finally, each element of $H$ possesses an inverse, because the inverse of each $u_i$ is contained in $H$ itself. Therefore, $H$ forms a subgroup in $G$. Note that $H$ is generated by {\it finitely} multiplying the group entries. Until now, we have generated a set in $G$ and have proved that it forms a subgroup.

{\noindent \bf Lemma\,2:} Let $S$ be a connected topological space. If $X \subseteq S$ is both open and closed ({\it clopen}), then either
\begin{align} \label{tog3a}
X = \varnothing \,({\rm null\,set}) \,\,\,{\rm or}\,\,\, X=S.
\end{align}

{\noindent \bf Proof:} Let's assume that $X$ is nonempty and clopen, i.e., $X \neq \varnothing$. Then $S \setminus X$ is also closed (since $X$ is open) and open (since $X$ is closed). So $S \setminus X$ is also clopen. Also, $S=X \cup (S \setminus X)$, namely, $S$ is a union of two disjoint sets. However, this contradicts the assumption that $S$ is a connected space. Therefore, we require either $X = \varnothing$ or $X=S$. Now, we are equipped to prove the main theorem.

\begin{figure}[ht]
\centering
\includegraphics[scale=0.7]{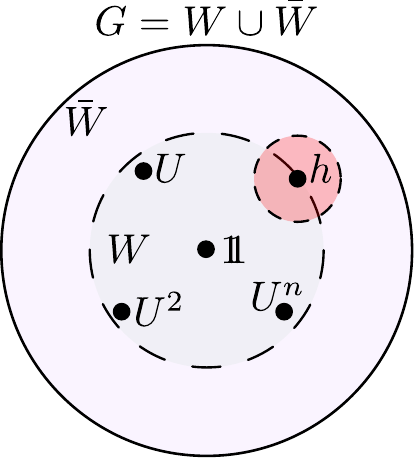}
\caption{Illustration of the topological proof of the theorem. We show that $G=W$, and $\bar{W}$ is a null set.}
\label{fig-cg}
\end{figure}

{\noindent \bf Theorem:} Let $G$ be a connected topological group, and let $U \subseteq G$ be an open neighborhood of the identity $\mathds{1}$. Let $H=\langle U \rangle$ be the subgroup generated by $U$. Then $H=G$. In other words, any neighborhood of the identity in a connected topological group generates the whole group.

\begin{figure}[htbp]
\centering
\includegraphics[scale=0.38]{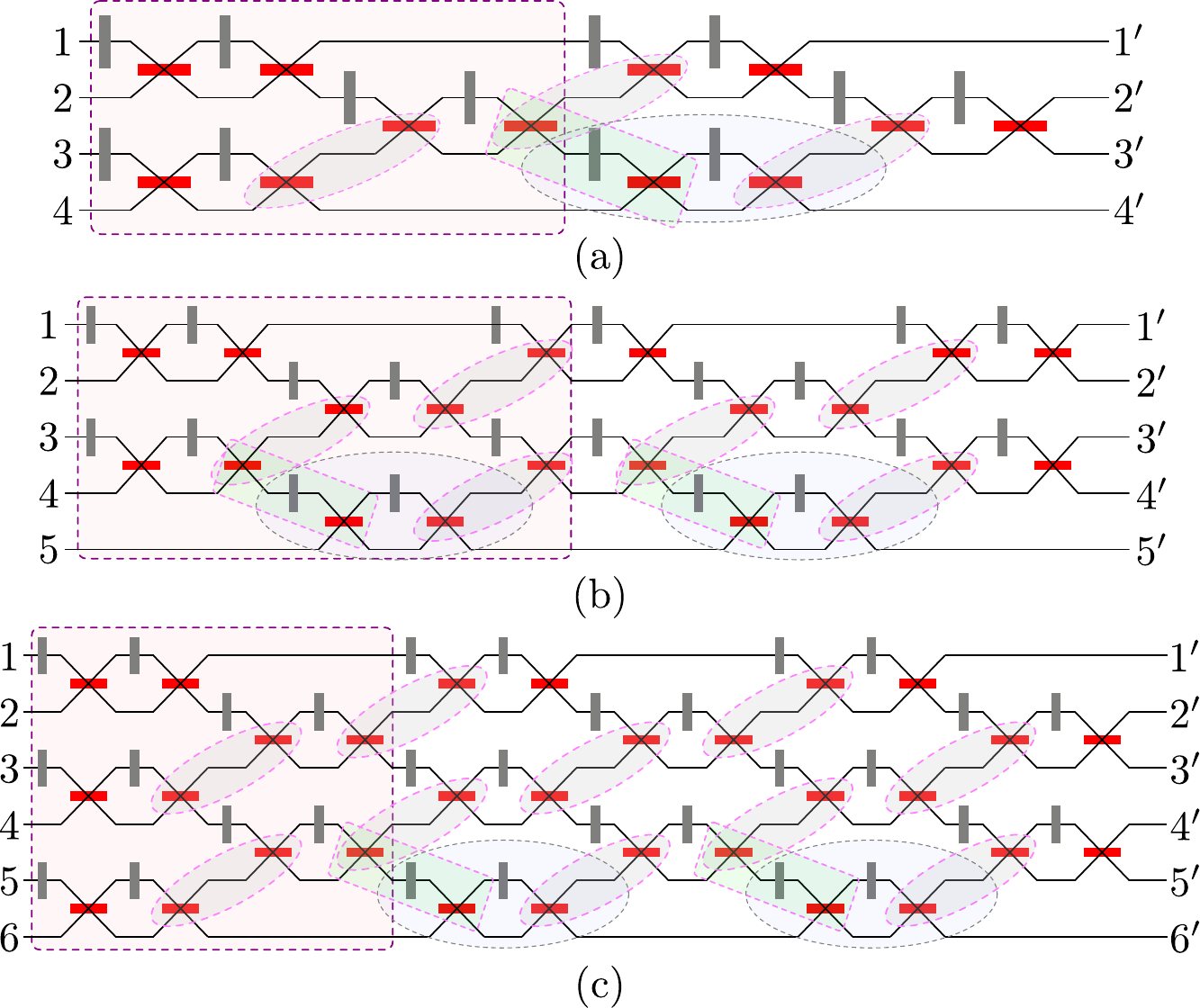}
\caption{Panels (a)-(c) show the Clements decompositions for dimensions 4, 5, and 6, excluding the diagonal matrix $D_N$ [see Eq. (\ref{lo2})]. The recurring rectangular blocks highlight the fundamental unit implied by the Lie group theorem. Each rotated elliptical region indicates that two BSs can be effectively replaced by a single fixed tritter, thereby reducing the insertion loss due to individual optical element. Moreover, substituting the aMZI unit\,(denoted by elliptical blocks) from Fig.~\ref{fig-mz}(a) with the modified version in Fig.~\ref{fig-mz}(b) allows further loss reduction, as adjacent BSs can be combined--illustrated by the rotated rectangles.}
\label{fig-6d}
\end{figure}

{\noindent \bf Proof\,1:} Since $U$ is an open neighborhood of $\mathds{1}$, the subgroup $H$, generated by the elements of $U$, is also open. If $H$ is open, then its cosets, denoted by $gH$, where $g \in G$, are also open. Now consider $X=\bigcup\limits_{g \in G} gH$, the union of all cosets of $H$. We know that $X$ is also open and is a complement of $H$, i.e., $X=G\setminus H$. Since $X$ is open, $H$ must be closed. Hence, $H$ is clopen. As $H$ is both clopen and a nonempty set and $G$ is connected, it should be just $G$. This completes the first proof.

{\noindent \bf Proof\,2:} Here we present a variant of the topological proof, following Ref.~\cite{pontrjagin46}. Let us assume that $U$ is an open neighborhood of $\mathds{1}$, the identity element of the group $G$. Because $U$ is open, we can also form another open set $W$ as
\begin{align} \label{pt1}
W = \mathds{1} \cup U \cup U^2 \cup \cdots \cup U^n.
\end{align}
Our aim is to prove that $W$ is a closed set. We note that $G=W \cup \bar{W}$, where $\bar{W}$ is the closure of $W$. Suppose there exists some $g \in \bar{W}$. Now $gU^{-1}$ intersects $W$, as it lies in the open neighborhood of $g$. In other words, we can find an element $h$ such that $h \in W \cap gU^{-1}$. Because $h \in W$, we can write $h=u_1 \ldots u_n$, where $u_1,\,\ldots,u_n \in U$. Because $h \in gU^{-1}$, we have $h=gu^{-1}$, with $u \in U$. Therefore, $g=u_1\ldots u_nu$, or equivalently, $g \in U$. Consequently, we find that $W$ is closed. We then conclude that $W=G$\,(or $\bar{W}$ is a null set), as $G$ is a connected group and $W$ is both open and closed\,(see Figure\,\ref{fig-cg}). That is, $G$ is completely generated by $U$. \hfill $\square$

When the PS elements are chosen appropriately, each MBS lies in a neighborhood of the identity, so a finite number of MBSs suffices to realize an arbitrary $U_N$. To see this, consider two MBSs, $u_1 = \exp(iH_1)$ and $u_2 = \exp(iH_2)$, where $H_1$ and $H_2$ are Hermitian generators of their respective unitary matrices. For example, the generators of $U(2)$ are the Pauli matrices, while those of $U(3)$ are the Gell-Mann matrices~\cite{georgi2000}. By the Baker-Campbell-Hausdorff (BCH) formula~\cite{hall2015}, their product $u_1u_2$ yields new effective Hermitian generators, and successive multiplications can generate the full Lie algebra of the unitary group. However, this depends on the choice of initial generators: for instance, starting with $H_1 = H_2 = \sigma_x$ as in Eq.\,(\ref{mz1}) fails to generate $\sigma_y$ and $\sigma_z$, so the full $U(2)$ cannot be realized. Referring to Figure\,\ref{fig-6d}\,(a), we observe that any element of ${\rm SU}(4)$ lies in $U \cup U^2$. Equivalently, any ${\rm SU}(4)$ can be written as $u_1$ or $u_1u_2$, where each $u_i$ denotes the recurring rectangular block, with the PS elements chosen arbitrarily.

\bibliography{Reference}% Produces the bibliography via BibTeX.

\end{document}